# Ductile Breakup of Tracer Aggregates in Homogenous Isotropic Turbulence

Graziano Frungieri[a,*], Matthaus U. Babler[b], Luca Biferale[c], Alessandra S. Lanotte[d]

[a]Department of Applied Science and Technology, Politecnico di Torino, Torino, Italy
[b]Department of Chemical Engineering, KTH Royal Institute of Technology, Stockholm, Sweden
[c]Department of Physics and INFN, University of Tor Vergata, Rome, Italy
[d]CNR NANOTEC and INFN, Sez. Lecce, Lecce, Italy
 graziano.frungieri@polito.it

In this paper we study the ductile breakup of tracer aggregates in an incompressible, homogeneous, and isotropic three-dimensional turbulent flow. The flow dynamics is studied by means of a direct numerical simulation, whereas the Lagrangian velocities and stress statistics along trajectories are obtained by particle tracking. We investigate the breakup dynamics under the hypothesis that aggregates are able to deform and accumulate energy. Within this framework, breakup occurs when the energy transferred to the aggregate by the flow exceeds a critical value. We contrast our predictions for ductile breakup with those obtained for brittle breakup. We observe that turbulence intermittency is crucial for the breakup of brittle aggregates, while it becomes less relevant for ductile aggregates. In the limit of highly ductile aggregates the breakup rate is dictated by the mean properties of the flow. We propose a simple model to capture this behaviour.

## 1. Introduction

The fragmentation of particle aggregates in a fluid flow is a phenomenon of broad interest in physical, chemical and environmental problems, including technological applications such as the processing of materials in the food, pharmaceutical and composite industry (Vasquez et al., 2022; Vasquez et al., 2023) or the formation and destruction of particles in the ocean (Andrady, 2017). For small aggregates, breakup is caused by the hydrodynamical shear stresses due to the flow motion, and traditionally it has been assumed to occur in a brittle manner, i.e., to occur instantaneously as soon as the aggregate happen to experience for the first time a fluid dynamic stress exceeding its internal strength (Frungieri et al., 2022).

However, depending on their internal structure and colloidal particle-particle interactions (Frungieri and Vanni, 2021), aggregates are expected also to be able to undergo ductile breakup, i.e., to store the energy transmitted by the fluid stress in internal deformation and to fail only when the accumulated energy exceeds their toughness limit (Marchioli and Soldati, 2015). Accumulation of energy transmitted to the aggregate structure through the hydrodynamic stress was considered by Saha et al. (2016) for the interpretation of experiments on the breakup of single aggregates in turbulence.

In either case, brittle or ductile breakup, a physical understanding of the process in complex flow conditions, such as those of turbulence, is still lacking, due to the difficulties of having at the same time a detailed description of the aggregate structure – counting for both hydrodynamic and colloidal interactions between constituent particles – and an accurate description of the flow dynamics (Brandt and Coletti, 2022; De Bona et al., 2014; Breuer and Khalifa, 2019).

In some studies, a simplified approach has been adopted, which consists in considering fully the complex turbulent flow dynamics, while drastically reducing the complexity of the aggregate structure, by considering it as a point-particle (Bäbler et al., 2012). By such an approach, the breakup of small, brittle aggregates (Bäbler et al., 2012; Bäbler et al., 2015) has been investigated in different flow configurations. In particular, the breakup rate was measured at varying strength of the aggregates, showing that the fragmentation mechanism has two distinct regimes. For loose aggregates, the fragmentation rate is high, and it has a universal power-law behaviour governed by the smooth, Gaussian fluctuations of the turbulence. For stronger aggregates, the rate of breakup is instead smaller, and its occurrence is controlled by the intermittent and intense burst of the turbulent stress.

Within an approach similar to the one used by Marchioli and Soldati (2015), in this work, we compute the breakup rate of small, ductile tracer aggregates, i.e., aggregates that follow passively the fluid streamlines and that break only when the accumulated energy overcome their toughness limit. To do this, we use data from a Direct Numerical Simulation of a three-dimensional isotropic turbulent flow at moderate Reynolds number, and we seed the flow with a large number of tracer aggregates. Two aggregate characteristic parameters (the critical stress for initiating the deformation process and the critical accumulated energy) are deemed as crucial and their effect on the breakup rates is investigated. Our interest in calculating breakup rates is motivated by the possibility offered by population balance models of accurately and efficiently tracking the evolution of the particle size distribution in process scale simulations (Lins et al., 2022; Frungieri and Briesen, 2023; Schiele et al., 2023). The paper is organised as follows: in Section 2, we report the equations used to describe the particle and flow dynamics and the approach used to track the accumulation of shear stresses on the aggregate structure; in Section 3 we discuss results for ductile breakup and we contrast them with those obtained for brittle aggregates, and with the predictions that can be obtained by simple modeling. Concluding remarks follow.

## 2. Methods

We consider a dilute suspension of aggregates described as point-like tracer particles, which have no feedback on the flow in which they are suspended, and which have no hydrodynamical interactions between them. Aggregates are smaller than the Kolmogorov scale of the flow $\eta$ and are treated as tracers carried passively by the flow. Their equation of motion thus reads as:

$$\dot{\mathbf{x}}_p = \mathbf{u}(\mathbf{x}_p, t) \tag{1}$$

where $\mathbf{x}_p$ is the particle position and $\mathbf{u}$ the fluid velocity. The latter was evolved according to the incompressible Navier-Stokes (NS) equations reading as:

$$\frac{\partial \mathbf{u}}{\partial t} + \mathbf{u} \cdot \nabla \mathbf{u} = -\frac{\nabla p}{\rho_f} + \nu \nabla^2 \mathbf{u} + \mathbf{F}, \qquad \nabla \cdot \mathbf{u} = 0. \tag{2}$$

where $\rho$ and $p$ are the fluid density and pressure, respectively, and where $\mathbf{F}$ is a forcing term injecting energy in the first low-wave number shells and keeping constant their spectral content (Bec et al., 2010). The NS equations are solved on a $512^3$ cubic grid with periodic boundary conditions, and a Taylor-scale Reynolds number $Re_\lambda \simeq 185$. The kinematic viscosity is chosen in such a way that the Kolmogorov length scale equals the grid spacing $\eta \simeq \delta x$. In Table 1 the main characteristics of the flow are reported. Further numerical details can be found in the work by Bec et al. (2010). The stress acting on the particles is the one due to shear only, which is computed along trajectories as (Kusters, 1991):

$$\sigma(\mathbf{x}_p, t) = \mu \sqrt{\frac{2}{15} \frac{\varepsilon(\mathbf{x}_p)}{\nu}} \tag{3}$$

where $\varepsilon(\mathbf{x}_p)$ is the local turbulent energy dissipation rate computed as $\varepsilon = 2\nu e_{ij} e_{ij}$ with $e_{ij}$ being the rate of deformation tensor, and where $\nu$ and $\mu$ are the kinematic and dynamic viscosity of the fluid, respectively.

We are interested in assessing the occurrence of ductile breakup. We assume that the breakup process has to be first activated (and this occurs when the hydrodynamic stress $\sigma$ acting on the aggregate exceeds a critical value $\sigma_{cr}$, that is a characteristic of the aggregate internal strength) and then it proceeds through the accumulation of energy until a critical threshold is reached. As long as the condition $\sigma > \sigma_{cr}$ is met, the aggregate stores energy as:

$$E(\tau) = \int_0^\tau \sigma(\mathbf{x}_p, t) \, \theta(\sigma - \sigma_{cr}) dt \tag{4}$$

where $\theta$ is the Heaviside step function. Breakup occurs when the accumulated energy exceeds a critical threshold $E_{cr}$. Hence, an individual aggregate that is released at a random time $t_0$ will break after a time-lag $\tau$ which is the time at which the accumulated energy, as computed from Eq.(4), assumes a value equal to $E_{cr}$ (that is a characteristic of the aggregate toughness limit). Figure 1 illustrates the approach just outlined. The breakup frequency follows as the inverse of the average time-lag obtained after tracking many aggregates. Formally, this can be written as:

$$f(\sigma_{cr}, E_{cr}) \equiv \frac{1}{\langle \tau(\sigma_{cr}, E_{cr}) \rangle}, \qquad \tau(\sigma_{cr}, E_{cr}) \equiv \left\{ \tau \;\middle|\; E_{cr} = \int_0^\tau \sigma(\mathbf{x}_p, t) \, \theta(\sigma - \sigma_{cr}) dt \right\} \tag{5}$$

where $f(\sigma_{cr}, E_{cr})$ is the breakup rate of aggregates characterized by $\sigma_{cr}$ and $E_{cr}$. Here $\tau(\sigma_{cr}, E_{cr})$ is the time-lag elapsed between the aggregate release in a flow region where $\sigma < \sigma_{cr}$ and the first time $E(\tau) = E_{cr}$. The brackets $\langle . \rangle$ indicate the ensemble average over the Lagrangian trajectories. We average the results over 128000 trajectories.

Table 1: Parameters of the DNS simulation: root-mean-square velocity $u_{rms}$, mean energy dissipation rate $\varepsilon$, kinematic viscosity $\nu$, Kolmogorov scale $\eta = (\nu^3/\langle\varepsilon\rangle)^{1/4}$, integral scale L, Eulerian large-eddy turnover time $T_E = L/u_{rms}$, Kolmogorov timescale $\tau_\eta = (\nu/\langle\varepsilon\rangle)^{1/2}$, $t_{run}$ simulation time.

| $U_{rms}$ | $\varepsilon$ | $\nu$ | $\eta$ | L | $T_E$ | $\tau_\eta$ | $t_{run}$ |
|---|---|---|---|---|---|---|---|
| 1.4 | 0.94 | 0.00205 | 0.010 | π | 2.2 | 0.047 | 13.2 |

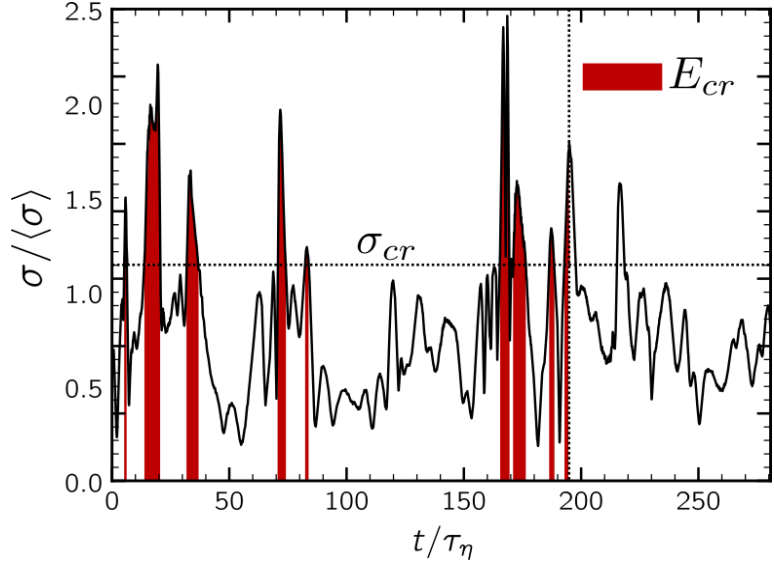

Figure 1: Illustration of the approach used to assess the occurrence of ductile breakup. Accumulation starts after the aggregate experience for the first time a shear stress $\sigma > \sigma_{cr}$. Breakup occurs when the energy accumulated exceeds the toughness limit $E_{cr}$. In the above example this happens at the time $t/\tau_\eta \cong 195$ indicated by the vertical dotted line.

In the limit of $E_{cr}=0$, i.e., for brittle aggregates, breakup occurs when the aggregate released at $t_0$ experiences for the first time a hydrodynamic stress that exceeds the critical stress $\sigma_{cr}$. This limiting case was investigated by Babler et al. (2012) who also provided a closed-form approximation of the breakup rate for brittle aggregates, based on an earlier model by Loginov (1986):

$$\tilde{f} = \frac{\int_0^\infty d\dot\sigma \, \dot\sigma \, p_2(\sigma_{cr}, \dot\sigma)}{\int_0^{\sigma_{cr}} d\sigma \, p(\sigma)} \quad (6)$$

In this expression, $p_2(\sigma, \dot\sigma)$ is the joint probability density function (PDF) of the hydrodynamic stress $\sigma$ and of its time derivative $\dot\sigma$ along the aggregate trajectory, and $p(\sigma)$ is the marginal PDF of the stress $\sigma$. Both $p_2(\sigma, \dot\sigma)$ and $p(\sigma)$ are computed along aggregate trajectories obtained by DNS.

## 3. Results

We compute first the probability density function of the energy that is accumulated by the aggregates over the whole length of their trajectories. We do this by assuming the aggregates to start accumulating energy as soon as they experience for the first time a stress larger than a critical threshold $\sigma_{cr}$, and by considering them as infinitely strong, i.e. resistant to breakup. In Figure 2 the results of the analysis are reported at varying values of the critical threshold $\sigma_{cr}$. For a low threshold of the critical hydrodynamic stress, all aggregates accumulate energy over nearly the entire length of their trajectory, leading to a relatively narrow PDF (black and orange curves in Figure 2. On the other hand, if the critical threshold is high, energy is accumulated only along the few segments of the trajectory where $\sigma > \sigma_{cr}$. Due to the turbulent flow variability, the local value of $\sigma$ and the length of these segments are strongly varying quantities, leading to a wider PDF (grey curve). The inset of Figure 2 shows the average of the accumulated energy, which decreases for increasing thresholds.

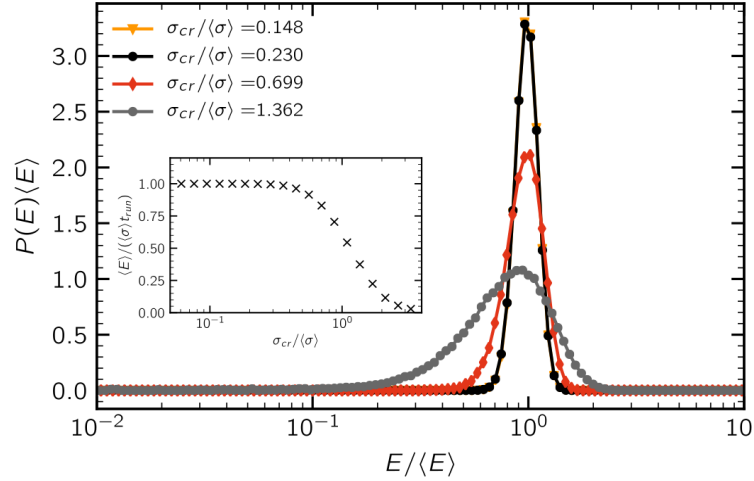

*Figure 2: Probability distribution function of the energy E accumulated by aggregates at varying value of the activation stress σ$_{cr}$. The critical stress has been made dimensionless by the average stress ⟨σ⟩. In the inset, the average energy is plotted as a function of the critical stress and normalized by the product between the average stress and the simulation run time.*

Figure 3 shows the breakup rate for various values of the accumulation energy $E_{cr}$ needed for breakup. The dashed line refers to the case of zero energy, i.e., is the one of brittle aggregates. For these, breakup occurs as soon as they experience the critical stress for the first time. Accordingly, the time lag for breakup is comparably short, and the breakup rate is high. As the threshold for the critical energy increases (i.e., as the aggregates become more tenacious), a longer stage of energy accumulation is necessary, and the breakup rate is lower. However, for both the brittle and ductile cases, when the critical stress is large, events where σ is larger than σ$_{cr}$ become rare; consequently, the breakup rate shows a rapid fall off.

When the breakup energy $E_{cr}$ is large, the aggregates spend a long time in the flow accumulating energy (long compared to the large eddy turn-over time $T_L$), and during this phase, they sample the whole stress probability space. We are willing to use this consideration as the basis for a model to describe the breakup rate: computing the accumulated energy as:

$$E_{cr} \simeq \langle \tau \rangle \int_{\sigma_{cr}}^{\infty} \sigma' p(\sigma') d\sigma' \tag{7}$$

where ⟨τ⟩ is the time lag for breakup (much larger than $T_L$ and of comparable duration among the different aggregates), we can solve Eq(7) for ⟨τ⟩ and evaluate the breakup rate as:

$$f(\sigma_{cr}, E_{cr}) = \frac{1}{\langle \tau \rangle} \simeq \frac{\int_{\sigma_{cr}}^{\infty} \sigma' p(\sigma') d\sigma'}{E_{cr}} \tag{8}$$

where p(σ) is the PDF of the shear stress, that is plotted in Figure 3b, whereas the prediction of Eq(8) is reported for the highest energy level considered in our simulations by the solid line in Figure 3a. The model correctly predicts both the plateau of the breakup rate at small critical stress and the fall off at larger critical stress. Deviations can be explained as the trajectories do not sample the whole probability space of the stress.

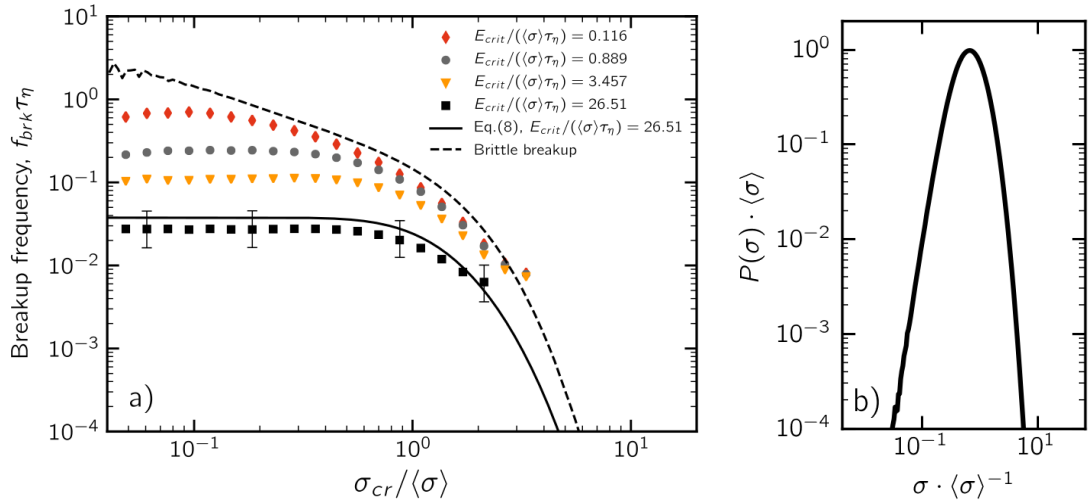

*Figure 3: a) Breakup frequency as a function of the critical activation stress. The breakup frequency is made dimensionless by the Kolmogorov length scale of the flow, whereas the critical stress is normalized with the average stress in the flow. Each data series (symbols) refers to a different value of the critical energy. The dashed curve refers to brittle particles. Eq(8) is plotted as a solid curve for the largest energy level investigated. b) Probability density function of the shear stress.*

### 4. Conclusions

In this work we have studied the ductile breakup of small aggregates in a homogenous isotropic turbulence by direct numerical simulations. We have treated aggregates as inertialess, tracer point-particles and we have tracked the history of shear stress they experience in the flow. We are interested in evaluating breakup rates. We have investigated the scenario in which aggregates are ductile, i.e., they undergo breakup only if they experience, at least once, a stress larger than a critical one (which can be thought of as the elastic limit upon which shear stresses induce irreversible deformation) and if the energy accumulated along the trajectory exceeds a critical energy threshold (which can be thought of as the aggregate toughness limit). Under these modeling conditions, at vanishing energy threshold, the usual mechanism for brittle breakup is recovered.

We have observed that for large activation stresses, the rate of breakup is controlled by the turbulence dynamics, and by the occurrence of the bursts of the turbulent hydrodynamic stress. On the other hand, for small activation stresses, turbulence fluctuations play a minor role: aggregates constantly accumulate stress along their trajectory and the breakup is independent of the dynamics of the stress and of the occurrence of intense turbulent bursts.

We have also observed that when aggregates have a large toughness limit, i.e., when they have to accumulate large energies in order to break, the contribution of the turbulent bursts of hydrodynamic stress become less relevant, and the occurrence of breakup can be predicted by simple modeling based on the average properties of the flow (Conchúir and Zaccone, 2013). On the contrary, for brittle aggregates, breakup is controlled by turbulent intermittency and occurs at large rates. Finally, our results confirm what was found by Marchioli and Soldati (2015) for the breakup of ductile aggregates in a bounded flow.

Future efforts could explore the use of breakup rates in population balance models to address the fragmentation dynamics and the evolution of the particle size distribution, also possibly in the presence of concurring aggregation phenomena.

**Acknowledgments**

M.U.B. acknowledges financial support from the Swedish Energy agency (Project Nr. P2019-90227). L. B. received funding from the European Research Council (ERC) under the European Union's Horizon 2020 research and innovation programme (grant agreement No 882340).